\documentclass[12pt,a4paper]{article}

\usepackage{cite}

\usepackage[onehalfspacing]{setspace}
\usepackage[left=2cm,top=2cm,right=2cm,bottom=2cm,nohead]{geometry}
\usepackage[format=plain,margin=0.5cm,textformat=simple,justification=justified,small,parskip=0cm]{caption}
\usepackage[pdftex]{graphicx}
\usepackage[pdftex,unicode,pdfstartview=FitH,colorlinks=true,pdfborder={0 0 0 0},urlcolor=blue]{hyperref}
\usepackage[cmex10]{amsmath}
\usepackage[caption=false,font=footnotesize]{subfig}

\newcommand{\rmd}{\mathrm{d}}

\usepackage{color}

\begin{document}

\title{Fluctuation analysis of the three agent groups\\herding model}

\author{Vygintas Gontis, Aleksejus Kononovicius}
\date{}

\maketitle

\begin{abstract}
We derive a system of stochastic differential equations simulating the dynamics of the three agent groups with
herding interaction. Proposed approach can be valuable in the modeling of the complex socio-economic
systems with similar composition of the agents. We demonstrate how the sophisticated statistical
features of the absolute return in the financial markets can be reproduced by extending the herding
interaction of the agents and introducing the third agent state. As well we consider possible
extension of proposed herding model introducing additional exogenous noise. Such consistent microscopic and macroscopic model
precisely reproduces empirical power law statistics of the return in the financial markets.
\end{abstract}

\section{Introduction}

The large number of actors (agents), the non-linear interactions between them and the feedback
of the macroscopic behavior of the system on the microscopic behavior of the agents are the essential
properties of the complex socio-economic systems. These properties lead to themacroscopic fluctuations,
characterized by the power-law distributions and the power-law autocorrelations \cite{Farmer2009Nature, Lux2009NaturePhys,Schinckus2013ConPhys}. The observed empirical properties of the fluctuations in the complex socio-economic
systems are both important for the estimation of risks and also give essential information about the system.
This information allows us to develop various imitation models of the complex socio-economic systems, enabling the forecast
and the control of their behavior \cite{Kononovicius2012IntSys}.

We investigate models encompassing both microscopic description of the complex socio-economic
systems, using Markov jumps between different agent groups (states), as well as consistent macroscopic
description, using the non-linear stochastic differential equations (abbr. SDEs) obtained analytically from
the master equation \cite{Kononovicius2012PhysA,Kononovicius2012IntSys}.

In this contribution we consider the statistical properties of the fluctuations
in the three agent groups herding model \cite{Kononovicius2013EPL} perturbed by exogenous information flow noise.
We also compare the properties of the noisy model with the statistical properties of the empirical
data extracted from the NYSE Trades and Quotes database.

We start by generalizing the Kirman's herding model \cite{Kirman1993QJE} by introducing a variable inter-event times. Next we define the herding interaction between three agent groups. The three groups model is simplified by relating it to the financial markets. Further we couple the endogenous fluctuations of the agent system with the exogenous information flow noise. Finally we discuss the obtained results in the context of the proposed double stochastic model of the returns in the financial markets \cite{Gontis2010PhysA,Gontis2010Intech}.

\section{Generalization of the Kirman's herding model}

In \cite{Kirman1993QJE} Kirman noticed that entomologists and economists observe similar behavior in distinct systems.
For a system with constant number of agents $N$, having two choices, $1$ or $2$, with number of agents $X$ in state $2$ and consequently with the number $(N-X)$ of agents in state $1$,  Kirman proposed a Markovian chain with the following per-agent transition rates:
\begin{equation}
\eta_1(x,N)=\sigma_1+N h x, \quad \eta_2(x,N)=\sigma_2+N h (1-x),\label{eq:Kirman}
\end{equation}
where $x=\frac{X}{N}$, $h$ terms define the herding behavior, while $\sigma_i$ terms describe the rates of the individual decisions to change opinion.

In previous work \cite{Kononovicius2012PhysA} we proposed a generalization to the Kirman model by introducing the feedback of the macroscopic state, $x$, on microscopic transition rates accounting for a variable inter-event time $\tau(x)$. The generalized per-agent transition rates can be expressed as follows:
\begin{equation}
\eta_1(x,N)=\sigma_1+\frac{N h x}{\tau(x)} , \quad \eta_2(x,N)=\frac{\sigma_2 + N h (1-x)}{\tau(x)},\label{eq:KirmanTau}
\end{equation}
Then the macroscopic SDE of herding model with variable rate of herding interaction can be written as:
\begin{equation}
\rmd x = \left[ \varepsilon_1 (1-x) - \frac{\varepsilon_2 x}{\tau(x)} \right] \rmd t_s +
\sqrt{\frac{2 x (1-x)}{\tau(x)}} \rmd W_s,\label{eq:SDETau}
\end{equation}
where we introduced the time scaling $t_s=h t$ with new parameters  $\varepsilon_1=\frac{\sigma_1}{h}$ and $\varepsilon_2=\frac{\sigma_2}{h}$.

In \cite{Kononovicius2012PhysA,Ruseckas2011EPL} we have shown that non-linear transformation of variables $y=\frac{x}{1-x}$ (here $x$ is driven by eq. \eqref{eq:SDETau}) gives SDE:
\begin{equation}
\rmd y = \left[ \varepsilon_1 + y \frac{2-\varepsilon_2}{\tau(y)} \right] (1+y) \rmd t_s + \sqrt{\frac{2 y}{\tau(y)}} (1+y) \rmd W_s . \label{eq:ysdefull}
\end{equation}
If $\tau(y) = y^{-\alpha}$ and in the limit $y \gg 1$ we can consider only the highest powers of $y$ present in the SDE above, in such case the above SDE belongs to the general class of SDEs,
\begin{equation}
\rmd x = \left( \eta - \frac{\lambda}{2} \right) x^{2 \eta -1} \rmd t_s + x^\eta \rmd W_s . \label{eq:gensde}
\end{equation}
The above general class of SDEs is known to generate power-law statistics \cite{Ruseckas2010PhysRevE}. The stationary probability density function (abbr. PDF), $p(x)$, and power spectral density (abbr. PSD), $S(f)$, of the general class of SDEs are given by:
\begin{eqnarray}
& p(x) \sim x^{-\lambda}, \\
& S(f) \sim 1/f^\beta,\quad \beta = 1 + \frac{\lambda-3}{2 (\eta -1)}.
\end{eqnarray}
The parameters, of eqs. \eqref{eq:ysdefull} and \eqref{eq:gensde}, are related as follows $ \eta = \frac{3+\alpha}{2}$, $\lambda = \varepsilon_2 + \alpha + 1$.

Eq. \eqref{eq:gensde} was previously derived from the point processes and its ability to reproduce power-law statistics was grounded in \cite{Ruseckas2010PhysRevE}.
Many physical and social systems are characterized by the complex interactions among different components. The power-law autocorrelation in the output of these systems is a common characteristic feature \cite{Kobayashi1982BioMed, Ivanov1998EPL, Ashkenazy2001PhysRevLett, Ashkenazy2002PhysA, Ivanov2004PhysRevE, Podobnik2009PNAS}.
The applications of such stochastic model might include varying complex systems possessing power-law statistical features. The direct consequence of the comparison is the ability to control the power-law exponents, $\lambda$ and $\beta$, of the $y$ statistical features obtained from the agent-based model, eq. \eqref{eq:KirmanTau}, and its stochastic treatment, eq. \eqref{eq:ysdefull}. This can be used to reproduce $1/f^\beta$ noise with $0.5 < \beta <2$ (for details see \cite{Kononovicius2012PhysA,Ruseckas2011EPL}).

\section{Three state herding model}

We can extend the herding model by introducing the three state agent dynamics with the fractions of
agents in each state, $x_1$, $x_2$ and $x_3$. In this case there are six per-agent transition rates:
\begin{equation}
\eta_{ji} (x,N)= \sigma_{ji} + N h_{ji} x_i ,
\end{equation}
where $j$ is the index of the starting state, $i$ is the index for the destination state (i.e. agent leaves state $j$ to move to state $i$). As before, \cite{Kirman1993QJE,Kononovicius2012PhysA}, we assume that herding is symmetric, $h_{ij} = h_{ji}$.

Next we can use the one-step, or birth-death process, formalism (the formalism itself is discussed in \cite{VanKampen2007NorthHolland}, while the technical details are given in \cite{Kononovicius2013EPL}) to obtain the following Fokker-Plank equation for a time-dependent system state PDF, $\omega(x_1, x_2, t)$,
\begin{equation}
\partial_t \omega = -\sum_{i=1}^2 \partial_{x_i} \left[ D_i^1 \omega \right] + \sum_{i=1}^2 \sum_{j=1}^2 \partial_{x_i} \left\{ \partial_{x_j} \left[ D_{ij}^2 \omega \right] \right\}, \label{eq:fokplank}
\end{equation}
with
\begin{eqnarray}
& D_1^1 = \sigma_{21} x_2 + \sigma_{31} (1-x_2-x_1) - (\sigma_{12} + \sigma_{13}) x_1 , \quad D_2^1 = \sigma_{12} x_1 + \sigma_{32} (1-x_2-x_1) - (\sigma_{21} + \sigma_{23}) x_2 , \nonumber  \\
& D_{11}^2 \approx h_{12} x_1 x_2 + h_{13} x_1 (1-x_2-x_1)  , \quad D_{22}^2 \approx h_{12} x_1 x_2 + h_{23} x_2 (1-x_2-x_1) , \label{eq:fpterms} \\
& D_{12}^2 = D_{21}^2 \approx - h_{12} x_1 x_2 . \nonumber
\end{eqnarray}
We will achieve a considerable simplification of this approach after some
additional assumptions regarding financial market interpretation.

In the current agent-based modeling of financial markets the most common choice of three types of agents is:
fundamentalists, chartists optimists and chartists pessimists \cite{Cristelli2012Fermi}. Let us consider fluctuations
of market price $P(t)$ according its fundamental value $P_f$, based on fundamental knowledge of fundamentalists.
It is common to assume the excess demand of fundamentalists, $ED_f(t)$, as a given by \cite{Alfarano2005CompEco}
\begin{equation}
ED_f(t) = N_f(t) \ln \frac{P_f}{P(t)} = N_f(t) p(t) ,
\end{equation}
where $N_f(t)$ is a number of the fundamentalists inside the market. Such assumption ensures the long term convergence of
the market price towards its fundamental value $P_f$, here considered to be constant. $p(t)$ stands for the relative log-price, $p(t)=\ln \frac{P_f}{P(t)}$.

The pessimistic and optimistic chartists, are short-term traders, who estimate the future prices
based on its recent movement and external information flow. It is reasonable to assume that all chartists
at a given time are divided as optimistic, i.e., suggesting to buy and pessimistic, i.e., suggesting to sell.
The excess demand of the chartist traders, $ED_c(t)$, is given by \cite{Alfarano2005CompEco}
\begin{equation}
ED_c(t) = \bar{r}_0 [N_o(t) - N_p(t)] ,
\end{equation}
where $\bar{r}_0$ is a relative impact factor of the chartist trader, $N_o$ and $N_p$ are the total numbers of optimists and pessimists respectively.

As a market price is assumed to reflect the current supply and demand, the Walrasian scenario in its contemporary form may be expressed as
\begin{equation}
\frac{1}{\beta N} \partial_t p(t) = - n_f(t) p(t) + \bar{r}_0 [n_o(t) - n_p(t)] ,
\end{equation}
here $\beta$ is a speed of the price adjustment, $N$ a total number of traders in the market and $n_i(t) = \frac{N_i(t)}{N}$. By assuming that the number of traders in the market is large, $N \rightarrow \infty$, the expression for the relative log-price is obtained
\begin{equation}
p(t) = \bar{r}_0 \frac{n_o(t) - n_p(t)}{n_f(t)}. \label{eq:logprice}
\end{equation}
Consequently the return, $r(t)$, in the selected time window $T$ is given by
\begin{equation}
r(t) = \bar{r}_0 \left[ \frac{n_o(t) - n_p(t)}{n_f(t)} - \frac{n_o(t-T) - n_p(t-T)}{n_f(t-T)} \right] . \label{eq:retfull}
\end{equation}

Now let us make simplifications to the three group agent-based model. First of all we relate the states' population fractions to the described agent types:
\begin{equation}
x_1 = n_f , \quad x_2 = n_p , \quad x_3 = n_o .
\end{equation}
Now let us note that the optimism and pessimism are essentially identical:
\begin{equation}
\sigma_{23} = \sigma_{32} = \sigma_{cc} , \quad \sigma_{12} = \sigma_{13} = \sigma_{fc}/2 , \sigma_{21} = \sigma_{31} = \sigma_{cf}, \quad h_{12} = h_{13} = h_1 .
\end{equation}
It is also reasonable to assume that chartists among themselves interact $H$ times faster than with the fundamentalists, i.e.,
\begin{equation}
h_{23} = H h_1 , \quad H \gg 1, \quad \sigma_{cc} \gg \sigma_{cf}, \quad \sigma_{cc} \gg \sigma_{fc} .
\end{equation}

After some more technical mathematical steps with the eq. \eqref{eq:fpterms}, for details see \cite{Kononovicius2013EPL}, we derive a system of SDEs corresponding to the Fokker-Plank equation \eqref{eq:fokplank},
\begin{eqnarray}
& \rmd n_f = \left[ (1-n_f) \sigma_{cf} - n_f \sigma_{fc} \right] \rmd t + \sqrt{2 h_1 n_f (1-n_f)} \rmd W_1 , \\
& \rmd n_p = (1-n_f - 2 n_p) \sigma_{cc} \rmd t + \sqrt{2 H h_1 n_p (1-n_f-n_p)} \rmd W_2 .
\label{eq:SDEfp}
\end{eqnarray}
The derived SDEs are interdependent, while it would be preferable to have a system of
independent SDEs. Introducing a mood, $\xi(t) = \frac{n_o(t)-n_p(t)}{n_o(t)+n_p(t)}$, as a new variable instead of $n_p$, helps us to arrive at the independent equations.

In the final version of equations we scale the time, $t_s = h_1 t$, and appropriately redefine the model parameters: $\varepsilon_{cf} = \sigma_{cf} / h_1$, $\varepsilon_{fc} = \sigma_{fc} / h_1$, $\varepsilon_{cc} = \sigma_{cc} / (H h_1)$. At the same time we recall the generalization of the herding model, eqs. \eqref{eq:KirmanTau} and \eqref{eq:SDETau}, by introducing the additional variability of the event rate, $\frac{1}{\tau(n_f,\xi)}$. In case of the three group model of financial market one will get:
\begin{eqnarray}
& \rmd n_f = \left[ \frac{(1-n_f) \varepsilon_{cf}}{\tau(n_f,\xi)} - n_f \varepsilon_{fc} \right] \rmd t_s + \sqrt{\frac{2 n_f (1-n_f)}{\tau(n_f,\xi)}} \rmd W_{s,1} , \label{eq:nftau}\\
& \rmd \xi = - \frac{2 H \varepsilon_{cc} \xi}{\tau(n_f,\xi)} \rmd t + \sqrt{\frac{2 H (1-\xi^2)}{\tau(n_f,\xi)}} \rmd W_{s,2} , \label{eq:xitau}\\
& \tau(n_f,\xi) = \left[ 1 + \left| \frac{1-n_f}{n_f} \xi \right|^\alpha \right]^{-1} . \label{eq:taunfxi}
\end{eqnarray}
Note that in this setup $\tau(n_f,\xi)$ is defined, eq. \eqref{eq:taunfxi}, securing zero fluctuations, when number of chartists vanishes.
In \cite{Kononovicius2013EPL} we have shown that this model possesses a fractured PSD similar to the one introduced earlier in double stochastic model considered in \cite{Gontis2010PhysA, Gontis2010Intech}.

Described model contains two shortcomings. First of all, as we noticed in \cite{Kononovicius2013EPL}, the exponent of PSD for the absolute return are too high in the comparison with the empirical data. Also the model itself accounts only for the endogenous fluctuations of agents, when the external noise of information flow has to be accounted for, as well. Here we propose a very simple approach to integrate endogenous and exogenous fluctuations.

\section{Exogenous information flow noise}

It is widely accepted to describe the movements of stock price, $S(t)$, as a geometric Brownian process
\begin{equation}
\rmd S = \mu S \rmd t + \sigma(t) S \rmd W. \label{eq:GBprice}
\end{equation}
In the above $W$ is considered to be an external information flow noise and $\sigma(t)$ accounts for the stochastic volatility conditioned by the macroscopic state of the agent system. Here we consider only the most simple case, when $\sigma(t)$ fluctuations are slow in comparison with external noise $W$. In such case the return, $r_t (T)$, in the time period $T$ can be written as a solution of eq. \eqref{eq:GBprice}
\begin{equation}
r_t (T) = \left(\mu-\frac{1}{2} \sigma^2 \right) T + \sigma W(T).\label{eq:GBreturn}
\end{equation}
This equation defines instantaneous return fluctuations as a Gaussian random variable with mean $(\mu-\frac{1}{2} \sigma^2) T$ and variance $\sigma^2 T$.

In \cite{Gontis2010Intech,Gontis2010PhysA}, while relying on the empirical analysis, we have assumed that the return, $r_t (T)$, fluctuates as instantaneous q-Gaussian noise $\xi[r_0(x),\lambda]$ with $\lambda=5$ and driven by some stochastic process $x(t)$. The function $r_0(x)$ has a linear form
\begin{equation}
r_0(x) = b + a |x|, \label{eq:r0fun}
\end{equation}
where parameter $b$ serves as a time scale of exogenous noise and $\frac{b}{a}$ quantifies the relative input of exogenous noise. Proposed three state herding model and its combination with external noise driven price, eq. \eqref{eq:GBprice}, gives a new interpretation to the such approach. One just needs to replace the instantaneous Gaussian fluctuations of the return, eq. \eqref{eq:GBreturn}, by q-Gaussian noise. The $r_0(x)$ in the new interpretation should be a function of the log-price, $p(t)$, defined in eq. \eqref{eq:logprice} to model joint exogenous and endogenous fluctuations of the return,
\begin{eqnarray}
& r_t(T) =  \xi \left\{ r_0\left[ \mathrm{MA}(p(t),T) \right] \sqrt{T}, \lambda \right\} ,\label{eq:returnfinal} \\
& \mathrm{MA}(x(t),T) = \frac{1}{T} \int\limits_{t-T}^{t} x(s) \rmd s .
\end{eqnarray}
Eq. \eqref{eq:returnfinal} finally defines double stochastic model of return in financial markets, which integrates the endogenous herding fluctuations of the three groups agent system with exogenous noise of information flow here described by the instantaneous q-Gaussian  fluctuations.

\begin{figure}[!t]
\centering
\includegraphics[width=0.4\textwidth]{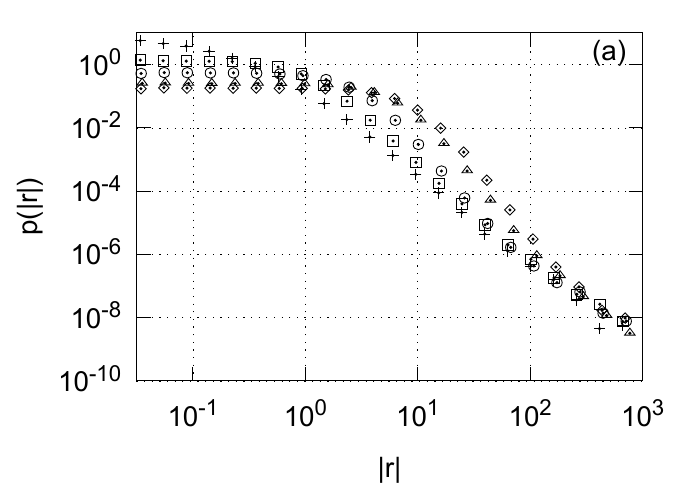}
\hspace{0.1\textwidth}
\includegraphics[width=0.4\textwidth]{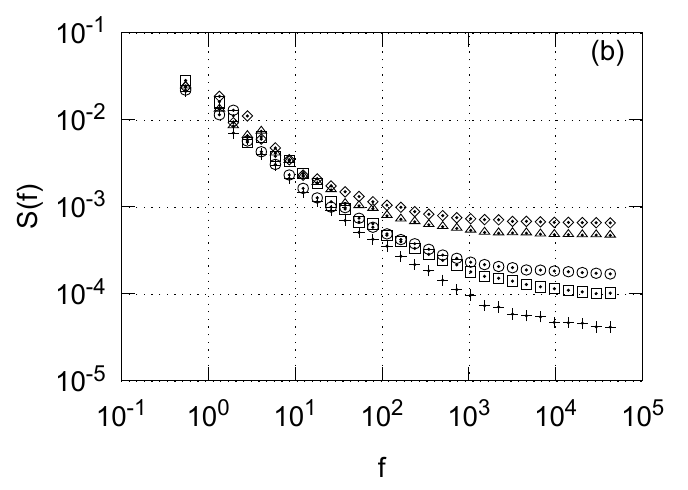}
\caption{The PDF (a) and PSD (b) of the absolute return, $|r|$, in the noisy three group model, eq. \eqref{eq:returnfinal}. The $q$-Gaussian noise here is with $\lambda=5$, and relative contribution of exogenous noise versus endogenous one is quantified by $\frac{b}{a} = 0.1$ (plus signs), $1.0$ (squares), $3.0$ (circles), $7.0$ (triangles) and $10.0$ (diamonds). The other model parameters were set as follows: $\varepsilon_{cf}=0.1$, $\varepsilon_{fc}=2$, $\varepsilon_{cc}=3.5$, $H=10$.}
\label{fig:logprice}
\end{figure}

First of all in fig. \ref{fig:logprice} we present numerical results of the extended three state herding model with exogenous noise for the return defined in eq. \eqref{eq:returnfinal}. It is obvious that external noise increases the exponent of return PDF in its power-law part, (a), and decreases both exponents of PSD, (b). Consequently it has to be possible to adjust exponents of power-law statistics for financial variables with appropriate choice of endogenous and exogenous noise contributions.

\begin{figure}[!t]
\centering
\includegraphics[width=0.4\textwidth]{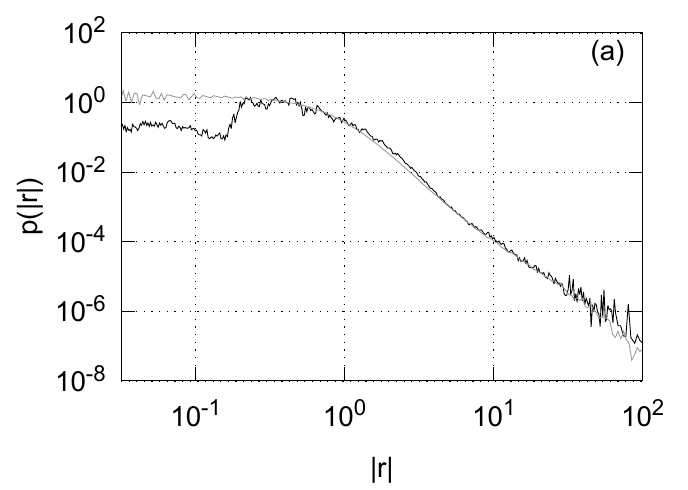}
\hspace{0.1\textwidth}
\includegraphics[width=0.4\textwidth]{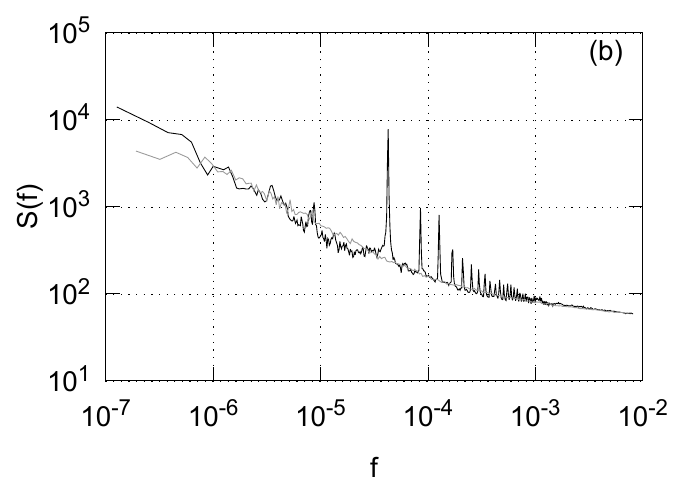}
\caption{Noisy three state model (gray curve) versus empirical data (black curve): PDF (a) and PSD (b). The $q$-Gaussian noise parameters were set as follows: $\lambda=5$, $a \sqrt{T} = 0.16$ and $b \sqrt{T}=0.9$. The other model parameters were set as follows: $\varepsilon_{cf}=0.5$, $\varepsilon_{fc}=2$, $\varepsilon_{cc}=3.5$, $H=10$, $h_1 = 1.66 \cdot 10^{-6} \mathrm{s^{-1}}$.}
\label{fig:empirical}
\end{figure}

In fig. \ref{fig:empirical} we compare the absolute return PDF, (a), and PSD, (b), numerically calculated with the noisy three state herding model and empirical data averaged over 24 series of different stocks traded on NYSE in two years period. This provides an evidence that proposed noisy three state herding model can reproduce empirical statistics of return in financial market in very details. From our point of view this result is considerable step in stochastic modeling of financial markets in comparison with previous modeling \cite{Gontis2010Intech,Gontis2010PhysA}, as incorporates microscopic model of agents and exogenous noise of information flow.

\section{Conclusion}
In this contribution we presented a possible extension of the three groups herding model recently proposed in \cite{Kononovicius2013EPL}. The main idea is
to incorporate into endogenous agent based model an external noise. In such approach the input of agent stochastic behavior and input of external noise into
common stochastic fluctuations are adjustable by parameters $a$ and $b$ in eq. \eqref{eq:r0fun}. By numerical calculations we demonstrate that exponents of power-law statistics, PDF and PSD, of long term return fluctuations are related to the choice of parameter $\frac{b}{a}$, figure \ref{fig:logprice}. This solves the shortcomings of earlier proposed three state herding model \cite{Kononovicius2013EPL}.

Finally we adjust parameters of extended three state herding model to reproduce statistics of return fluctuations in real markets, fig. \ref{fig:empirical}. Averaged PDF and PSD over 24 two years return series of stocks traded on NYSE are reproduced in very details by proposed model. Though this result resembles very much previous modeling of financial markets by SDEs \cite{Gontis2010Intech,Gontis2010PhysA}, current  approach is much more general as derived from microscopic treatment of complex social system. We do consider proposed model as potentially applicable to other social systems with similar three state composition of agents.

\bibliographystyle{ieeetr}
\bibliography{physrisk}

\end{document}